\documentclass{jaa}
\usepackage{natbib}
\usepackage{graphicx}

\begin{document}\sloppy

\title{New Results from the UVIT Survey of the Andromeda Galaxy}


\author{Leahy, D.A.\textsuperscript{1,*}, Postma, J.\textsuperscript{1}, Buick, M.\textsuperscript{1} 
 Morgan, C.\textsuperscript{1},  Bianchi, L.\textsuperscript{2}, and Hutchings, J.\textsuperscript{3}}
\affilOne{\textsuperscript{1}Department of Physics and Astronomy, University of Calgary, Calgary, Canada.\\}
\affilTwo{\textsuperscript{2}Johns Hopkins University, Baltimore, MD, USA.\\}
\affilThree{\textsuperscript{3}Herzberg Institute of Astrophysics,Victoria, BC, Canada.}


\twocolumn[{

\maketitle

\corres{leahy@ucalgary.ca}

\msinfo{31 Oct 2020}{}

\begin{abstract}
The Andromeda Galaxy (M31) has been observed with the UltraViolet Imaging Telescope (UVIT) 
instrument onboard the AstroSat Observatory.
The M31 sky area was covered with 19 fields, in multiple UV filters per field, over the period of 2017 to 2019. 
The entire galaxy was observed in the FUV F148W filter, and more than half observed in the NUV filters.
A new calibration and data processing is described which improves the astrometry and photometry of the UVIT data.
The high spatial resolution of UVIT ($\sim$1 arcsec) and new astrometry calibration ($\sim$0.2 arcsec) allow identification of 
UVIT sources with stars, star clusters, X-ray sources, and other source types within M31 to a much better level than previously possible. 
We present new results from matching UVIT sources with stars measured as part of the Pan-chromatic Hubble
Andromeda Treasury project in M31.
\end{abstract}


}]


\doinum{12.3456/s78910-011-012-3}
\artcitid{\#\#\#\#}
\volnum{000}
\year{0000}
\pgrange{1--}
\setcounter{page}{1}
\lp{9}

\section{Introduction}
M31 is the closest neighboring large galaxy to our Galaxy. It is a large spiral having many similarities to our Galaxy
and can be used as a template to study the many aspects of our Galaxy which are difficult to study because
of our location inside it and the resulting high extinction to much of our Galaxy.
Another advantage of studying objects in M31 is that it is at a well known distance (783 kpc, McConnachie et al. 2005)
thus the uncertainty in intrinsic brightness for many objects is better known than for most Galactic sources.

M31 has been observed in optical on numerous occasions. 
The highest resolution observations are carried
out with the Hubble Space Telescope, including the  Pan-chromatic Hubble Andromeda Treasury (PHAT)
survey (Williams et al. 2014). 
In near and far ultraviolet (NUV and FUV), the GALEX instrument (Martin et al. 2005) has surveyed M31.

AstroSat  has four instruments, covering NUV and FUV 
with the UltraViolet Imaging Telescope (UVIT), and soft through hard X-rays with the Soft X-ray Telescope (SXT), 
Large Area Proportional Counters (LAXPC) and Cadmium-Zinc-Telluride Imager (CZTI) instruments  (Singh et al. 2014).
We are carrying out a survey of M31 in NUV and FUV with UVIT. 
 
UVIT observations have high spatial resolution ($\simeq$1 arcsec) and have capability of 
resolving individual stellar clusters and a large number of individual stars in M31.
 Previous observations of M31
with UVIT were presented in part by Leahy et al. (2020a), Leahy \& Chen (2020) and  Leahy, Bianchi \& Postma (2017).
Those papers presented an M31 UVIT point source catalog, matching M31 UVIT sources with Chandra sources, and analysis of 
UV bright stars in the bulge, respectively.

In this paper we describe new UVIT data processing which gives important 
improvements in astrometric accuracy and photometric accuracy for  UVIT data.
We compare the UVIT Field 2 data to observations in optical from the PHAT survey to study
the properties of the stars northeast of the bulge of M31.

\begin{figure*}
\centering\includegraphics[height=.5\textheight]{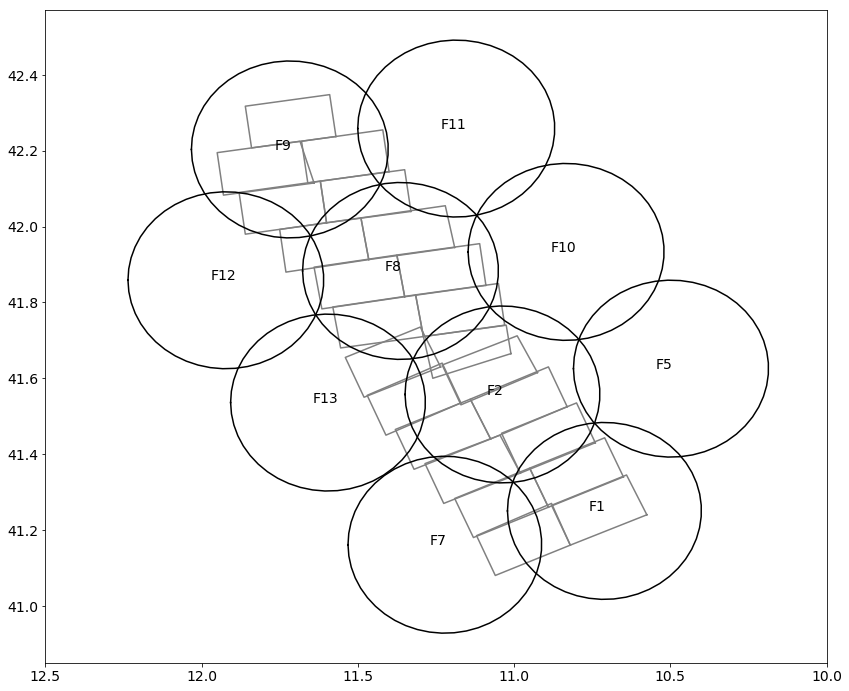}
\caption{Sky positions of the 10 fields (black circles) that make up the northern half of M31 UVIT survey, in J2000 coordinates. 
Overlaid on the UVIT fields are the outlines (grey rectangles) of the 23 areas  (`PHAT bricks') observed by HST in the PHAT project (Williams et al. 2014).}
\end{figure*}

\begin{table*}
\tabularfont
\caption{M31 UVIT Observations for Fields With N279N Data Overlapping the PHAT Survey}
\centering
\begin{tabular}{ccclccc}
\topline
Field & RA(deg)$^{1}$ & Dec(deg)$^{1}$ & Filter$^{2}$ & Exposure Time & Mean BJD$^{3}$ \\
\hline
\hline
1     & 10.71071 & 41.25023 & a, e, f & 7872, 7920, 4347 & 2457671 (+0.8010,+0.8010,+1.2626) \\
    2     & 11.03700 & 41.55735 & a, d, e, f & 7940, 16022, 7977, 7424 & 2457704 (+0.1355,+0.4858,+0.1354,+0.4857) \\
    7     & 11.22142 & 41.16111 & a, d, e, f & 4965, 10693, 10774, 3147 & 2458071 (+0.1875,+0.4520,+0.4520,+0.1874) \\
    13    & 11.59533 & 41.53595 & a, d, e, f & 4974, 10609, 10848, 5005 & 2458092 (+0.2157,+0.4981,+0.4327,+0.2157) \\
\hline
\end{tabular} 
\tablenotes{1: RA and Dec are the J2000 coordinates of the nominal pointing center of the observation.}
\tablenotes{2: Filter labels are a: F148W, b: F154W, c: F169M, d: F172M, e: N219M, f: N279N.}
\tablenotes{3: Mean BJD is the mean solar-system Barycentric Julian Date of the observation. The common integer part for multiple observations is given as the first number.}
\label{table:obs} 
\end{table*}

\section{Observations}  

UVIT consists of two 38 cm telescopes, each with field of view of $\sim$28 arcmin in diameter.
The UVIT telescope and calibration are described in Tandon et al. (2017a, 2017b, 2020), 
Postma et al. (2011),  Leahy et al. (2020b) and references therein.
 One telescope is for far ultraviolet (FUV) (130 to 180 nm) wavelengths and 
one for near ultraviolet (NUV) (200 to 300 nm) and visible (VIS) (320 to 550 nm) wavelengths.
The FUV, NUV and VIS channels each have a number of filters with different bandpasses.
The VIS channel is used for spacecraft pointing, so normally science observations are carried out simultaneously 
in FUV and NUV channels. 
The pixel scale for UVIT images is 0.4168 arcsec per pixel and 
point sources in the UVIT images have FWHM $\simeq$1 arcsec in the FUV and NUV channels. 

The survey of M31 with UVIT has been carried out since 2017.
With UVIT's 28 arcmin in diameter field of view, 19 different fields are required to cover the sky area of M31.
Exposures in the following filters are being used : F148W (123 to 173 nm), F154W (135 to 173 nm), 
F169M (146 to 175 nm), F172M (165 to 178 nm), N219M (206 to 233 nm) and N279N (275 to 284 nm). 
The NUV channel failed in early 2018, so that science observations since then have been carried out in FUV only.
Fortunately much of the UVIT survey which overlaps with the PHAT survey was observed prior to the NUV failure,
so that most of the PHAT area is covered by both NUV and FUV UVIT observations.
The sky positions of the 10 fields covering and adjacent to the area surveyed by HST as part of PHAT  
are shown in Fig. 1.

After the previous work (Leahy et al. 2020a) we developed new detector distortion corrections and new 
position calibration tools for processing UVIT images. 
Thus for the current work we are reprocessing the previous data on M31.  
For this study, we are comparing UVIT FUV and NUV data with NUV (the 275 nm band F275W filter) and optical data from the PHAT survey.
The UVIT N279N filter covers nearly the same waveband as the HST F275W filter.
Thus for the current analysis we restrict ourselves to the four UVIT Fields which overlap with PHAT (see Fig. 1) and also
have N279N data: Fields 1, 2, 7 and 13.
Table 1 here gives the basic properties these four Fields, including filters, exposure times and dates of observation.


\section{Data Analysis}

The UVIT detector distortion maps are utilized in the CCDLAB UVIT Pipeline (Postma 
\& Leahy 2017). We have modified the pipeline to reduce astrometric errors as well as a reduce the PSF of point sources. 
Previously the distorion maps were utilized simply at the unit pixel scale of the CMOS given that this is the scale at 
which they were measured upon, in comparison to the 1/8th pixel scale of the final science centroid-based images. 
And so although the distortion maps do not have appreciably high-frequency spatial components, there are 
nonetheless finite differentials in distortion from any given CMOS pixel to the next. 
Thus, bilinear interpolation of the distortion map is now implented and is applied to each centroid at the 1/32 pixel scale.

Photometric calibration improvements have also been applied with updated filter-wise flat fields. 
Previously the flats were only at the detector level (Tandon et al. (2017a, 2017b), but ongoing 
in-orbit calibration has allowed for second-order corrections to these flat fields to be developed 
for each filter (Tandon et al. 2020).

Additionally, the previous WCS solutions were typically solved with approximately only ten sources across the field. 
With the development of the trigonometric WCS auto-solving algorithm (Postma \& Leahy 2020) now implemented in 
CCDLAB, typically hundreds of sources spread across the field are now used for the plate solution. 
This provides for a more accurate average WCS solution across the field, with solution residuals commonly 
falling at 0.2 arcsec standard deviation.

\begin{figure}[!t]
\includegraphics[width=.99\columnwidth]{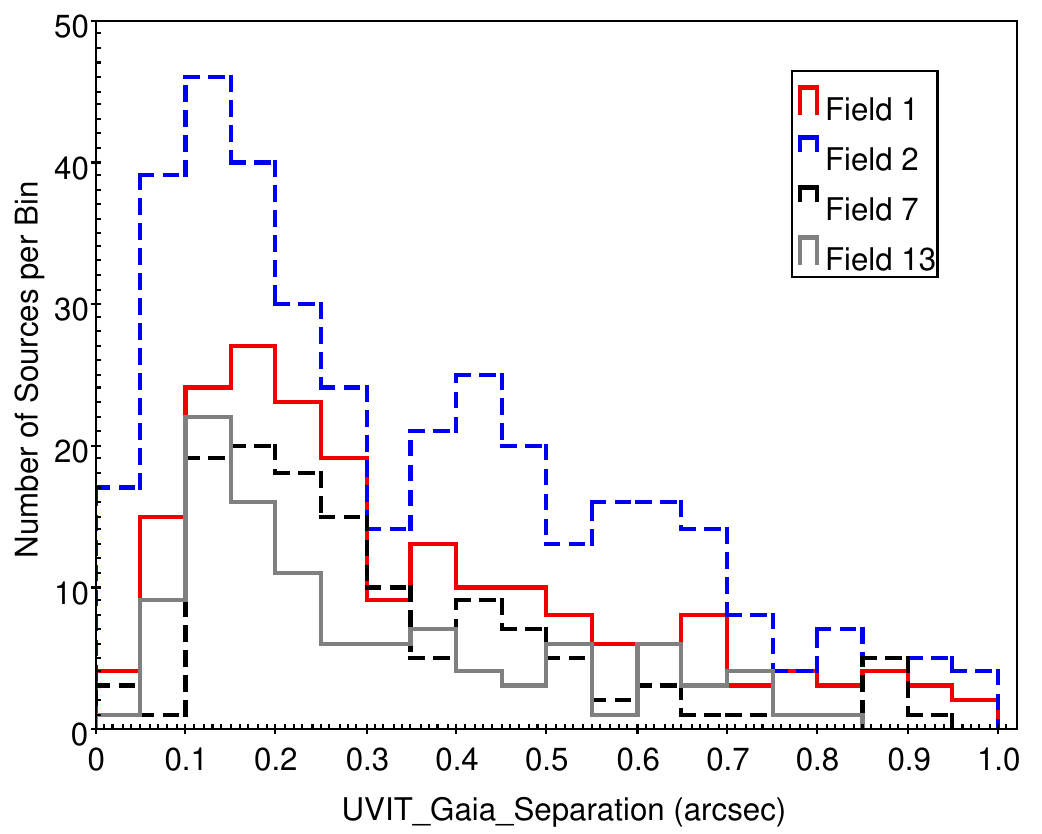}
\caption{ The distributions of Gaia-UVIT point source offsets for fields 1, 2, 7 and 13}\label{fig1}
\end{figure}

\begin{figure}[!t]
\includegraphics[width=.99\columnwidth]{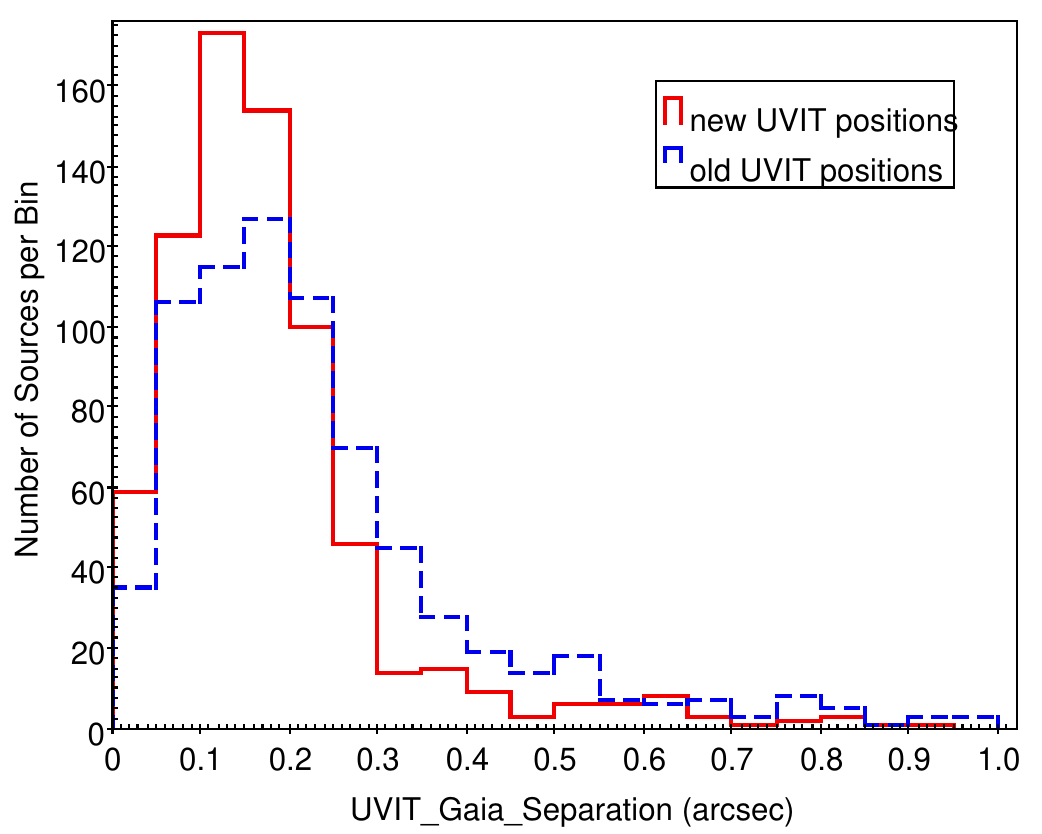}
\caption{ The distributions of the new-processing Gaia-UVIT offsets (red solid histogram) and the previous Gaia-UVIT offsets (blue dashed histogram)
 for point sources combined from Fields 1, 2, 7 and 13}\label{fig2}
\end{figure}

\begin{table}
\caption{Astronometry Errors for Fields Listed in Table 1}
\centering
\begin{tabular}{cccc}
\topline
Field & StDev$^{a}$ & No. Calibrators  \\
\hline
    1$^{b}$     & 0.192 & 271   \\ 
    2$^{b}$     & 0.177 & 215   \\ 
  7$^{b}$     & 0.182 & 273   \\
 13$^{b}$    & 0.176 & 332   \\
\hline
\end{tabular}
\footnotesize
$\quad$\\
a. StDev is the standard deviation of positions with respect to Gaia positions in units of arcsec. \\[0pt]
b. N279N is the filter image used for calibration with Gaia positions. \\[0pt]
\label{table:posnerr} 
\end{table}

\subsection{Point Source Analysis}


This study aims to obtain UVIT photometry for stars already identified in the PHAT survey. 
Thus we searched for an excess above local background,  in each UVIT reprocessed image, within 1$^{\prime\prime}$ 
of the positions of the known PHAT sources. 
To avoid confusion of identifying a single UVIT source with either neighboring UVIT sources or with multiple PHAT sources, we only searched at the positions of PHAT sources which had F275W (275 nm band) magnitudes,
and which were separated from eachother by more than 3.75$^{\prime\prime}$\footnote{Tests showed that separations 
of 9 pixels (3.75$^{\prime\prime}$) was enough to obtain good flux measurements for adjacent UVIT sources, and this same separation was more than enough to avoid confusion with nearby PHAT sources}.
We used the CCDLAB software (Postma \& Leahy 2017) 
with a threshold of 3 sigma excess for a single pixel in the box and 8 sigma above local background for the total counts
in the box. 
The centroids of the excesses were kept in a list of candidate point sources.
Then we perform a PSF (point spread function) fit to a 9 pixel x 9 pixel box surrounding the position of the source to
derive a more accurate position and total counts for each source.

The UVIT PSF has extended wings (Fig. 5 and Table 7 of Tandon et al. 2017b).
Thus we made measurements of the correction from 9x9 box net counts to larger (17x17 pixels) box counts using  
fits to a set of bright isolated  point sources in each field where we could use the larger box without interfering sources.
An additional small correction was added to correct for the counts in a 17x17 pixel box to total counts in a 
circle of radius 27 pixels using the data in Tandon et al. (2017b). 
This way counts for sources separated by more than a few pixels were determined with reasonable accuracy.
The counts and exposure times were used to convert to AB magnitudes using the updated calibration in Tandon et al. (2020).

The position uncertainties for UVIT were determined as follows.
The images were registered to optical position calibrators from Gaia as part of the reprocessing the images described
in the previous section above. 
The standard deviation of the resulting Gaia-UVIT offsets ranged from 0.17 arcsec to 0.19 arcsec 
(see Table 2).
Fig. 2 shows the distributions of all individual Gaia-UVIT offsets for fields 1, 2, 7 and 13 from the new data processing. 
Fig. 3 shows the distributions of the new-processing Gaia-UVIT offsets and the previous Gaia-UVIT offsets
 for the  point sources combined from Fields 1, 2, 7 and 13.
The new offsets are smaller than those from the previous (pre-2020) UVIT data processing procedure.

\section{Results and Discussion}

Our analysis above produced a catalog of UVIT AB magnitudes in the different observed UVIT filters (see
Table 1) for PHAT sources with measured F275W magnitudes, and with separations between PHAT sources
$>3.75^{\prime\prime}$. 
For these individual stellar sources, we have the UVIT AB magnitudes for the different observed UVIT filters 
(Table 1) and PHAT Vega magnitudes from the PHAT source catalog (Williams et al. 2014).
The UVIT bands range from F148W ($\simeq$150 nm) to N279N ( ($\simeq$280 nm),
and the PHAT bands range from F275W  ($\simeq$275 nm) to F160W ($\simeq$1600 nm).
With this combined photometry, we can study the nature of individual stars in M31.

\begin{figure*}
\centering\includegraphics[height=.7\textheight]{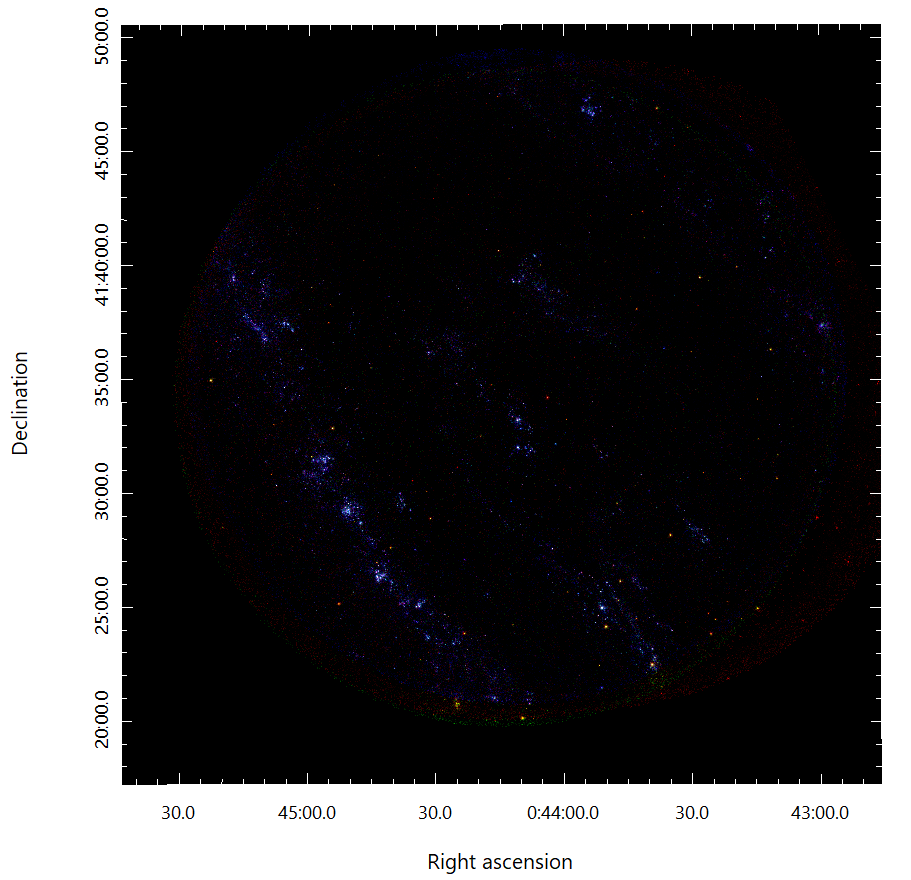}
\caption{ UVIT three-color image of the M31 survey Field 2, 28 arcmin in diameter. 
The N279N image is red, the N219M image is green, and the F148W image is blue. 
From edge to edge the image is $\simeq$6.4 kpc across at the distance of M31.
Coordinates are J2000 RA and Dec.}
\end{figure*}

The UVIT Field 2 has the best overlap with PHAT, so we concentrate the current work on Field 2.
A 3 color image was made using the 280 nm N279N (red), 220 nm N219M (green) and 150 nm F148W (blue)
images, which were position registered as part of the data processing.
Fig. 4 show the UVIT 3 color image of Field 2. The hot stars show up as blueish and the coolest stars as yellowish or
reddish. There are color artifacts around the edge of the image because we used full field images from all three 
filters, but the pointing centers for the three images were slightly different. 
This means that the outer edges of the images can have data from only one or two of the three filters, which results in
color errors, but not position distortion.

\begin{figure*}
\centering\includegraphics[height=.4\textheight]{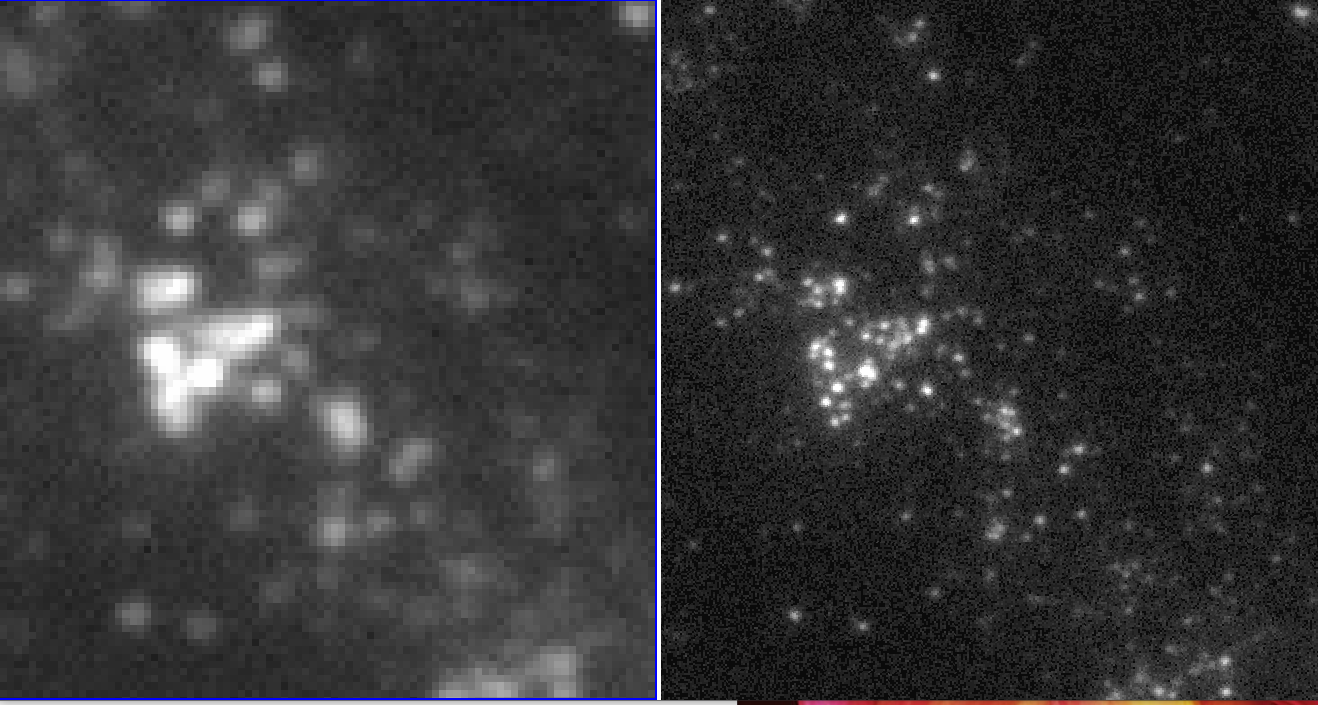}
\caption{Comparison of the GALEX image (left) with the UVIT F148W image (right) for the same region in
M31.  The area is 1.5 arcmin across and centered at R.A. 00:44:40.5 Dec.+41:26:37
(J2000).}
\end{figure*}

We compare the UVIT images, with $\sim$1 arcsecond resolution, to existing GALEX images of M31, which were taken 
with $\sim$5 arcsecond resolution. A concentration of hot stars is seen near RA 00:44:40, Dec. +41:26 
in the Field 2 three-color image in Figure 4.
In Figure 5 here we show an expanded view $\simeq$1.5 arcmin by 1.5 arcmin across centered at 00:44:40.5 Dec.+41:26:37
(J2000). The left hand panel is the deepest image from the GALEX survey (from the Deep Imaging Survey, DIS),
the right panel is the UVIT image in the F148W filter. The advantage of the higher resolution of UVIT is clearly seen,
as several blobs of emission in the GALEX image are resolved into separate sources in the UVIT image.
Here, the 3 brightest GALEX sources, just left of center in the GALEX image, are resolved into $\sim$19 separate 
sources in the UVIT F148W image. 

\begin{figure}[!t]
\includegraphics[width=.99\columnwidth]{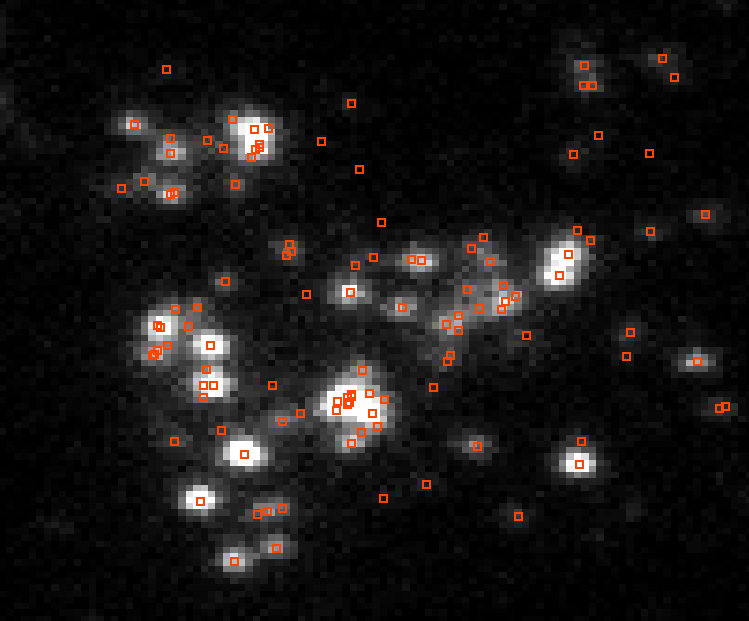}
\caption{Expanded view (greyscale) of the UVIT F148W image of Field 2, 30 arcsec across, 
centered on the bright group of stars  
east of R.A. 00:44:40.5 Dec.+41:26:37 (J2000) (center of the image in Fig. 3). 
The red squares mark the positions of all stars from the PHAT survey with F275W Vega magnitude $<20.5$.
}\label{fig2}
\end{figure}

Fig. 6 shows an expanded UVIT F148W image, 30 arcsec by 30 arcsec, of the concentration of stars to left of center in Figure 5.
UVIT is sensitive in N279N (same waveband as F275W) only to stars with N279N AB magnitudes brighter than $\simeq$22.
The conversion factor from AB to Vega magnitudes for the N279N filter is
$M_{AB}=M_{Vega}+1.48$ using the N279N filter response given in Tandon et al. (2017b).  
Thus this limit correspond to F275W Vega magnitudes brighter than $\simeq$20.5.
Overlaid on the UVIT image are the positions of all PHAT stars that have F275W Vega magnitude brighter than 20.5.
It is seen that all UVIT F148W (150 nm) detections have a F275W counterpart, but not all F275W stars are detected in F148W.
This is expected because only the hottest stars ($T_{eff}\gtrsim10,000$ K) are strong emitters for wavelengths as
short as 150 nm. 
In several locations multiple PHAT F275W stars overlap the UVIT F148W source. This occurs when there is source confusion
in the UVIT image. 
However, there are a significant number of UVIT F148W sources that can be identified with unique PHAT F275W sources.

\begin{figure}[!t]
\includegraphics[width=.99\columnwidth]{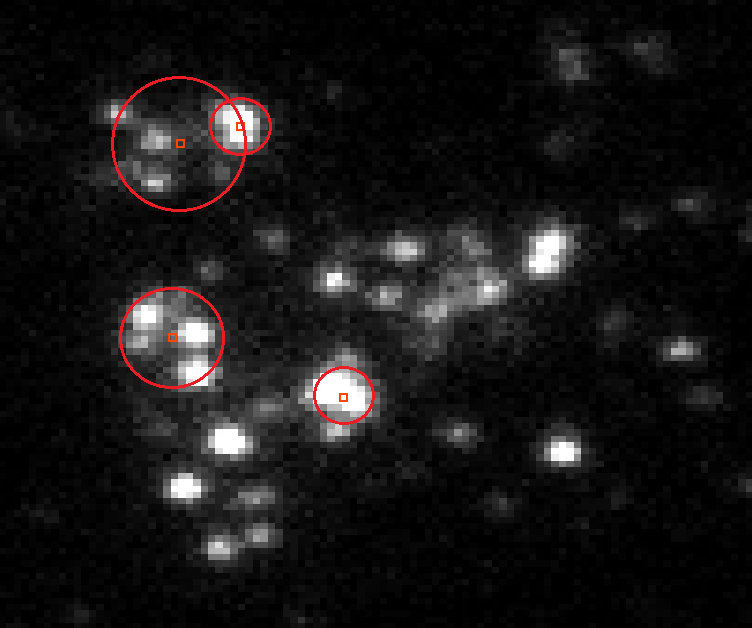}
\caption{Expanded view (greyscale) of the UVIT F148W image of Field 2, 30 arcsec across, 
centered on the bright group of stars  
east of R.A. 00:44:40.5 Dec.+41:26:37 (J2000) (center of the image in Fig. 3). 
The red circles with squares at their centers mark the stars clusters from Johnson et al. (2015)
that are in this region, with the radius of the circle set to the $R_{ap}$ from Johnson et al. (2015).
}\label{fig2}
\end{figure}

Figure 7 shows the same UVIT F148W image with the concentration of hot stars, but is now overlaid with
the stellar clusters listed in Johnson et al. (2015). Four of those clusters are located in the region, with centers
of the clusters marked by small red squares and with $R_{ap}$ of each cluster marked by a red circle.
Two of the clusters are compact ($R_{ap}$=2.66 and 2.63 arcseconds) and two are extended
 ($R_{ap}$=4.67 and 3.81 arcseconds). The compact clusters are not resolved by UVIT, and each
 compact cluster has several ($\sim$10) PHAT F275W stars in it (see Fig. 6).
 The extended clusters are mostly resolved by UVIT, with more than half of the UVIT sources in each cluster
 corresponding with single PHAT F275W stars (see Fig. 6). 
 Thus UVIT can resolve the larger clusters, $R_{ap}\gtrsim$3 arcseconds, in M31 into individual stars,
 and thus allow fitting photometry of individual stars.
 Most hot stars outside clusters are less crowded, so also can be identified with unique UVIT and PHAT identifications.

\begin{figure}[!t]
\includegraphics[width=.99\columnwidth]{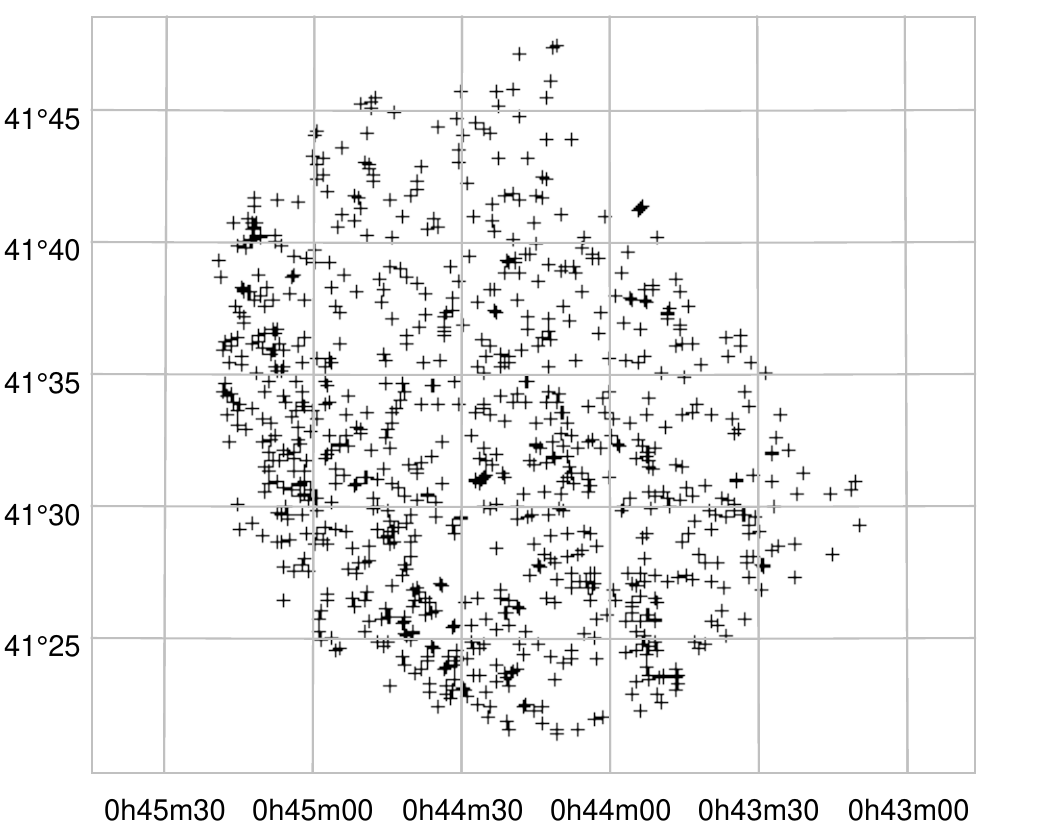}
\caption{Sky positions of the UVIT sources from Field 2 matched with unique PHAT sources.}\label{fig4} 
\end{figure}

For this work we consider the color-magnitude diagram (CMD) for UVIT sources which can be identified with individual 
PHAT stars. 
Model fitting of individual stars UVIT photometry combined with PHAT photometry will be done in future work.
To study the CMD, we use our list of all isolated UVIT sources with unique F275W PHAT counterparts, that are not also
associated with the known clusters in M31 from Johnson et al. (2015).

We found all UVIT sources with N279N detections which match with a PHAT F275W source within 1$^{\prime\prime}$ in Field 2,
then removed any UVIT sources  which are within 9 pixels ($\simeq 3.75^{\prime\prime}$) of a second PHAT F275W source.
Then we removed those sources with inconsistent N279N and F275W magnitudes, which indicates accidental matches.
\begin{table*}
\caption{Photometry$^{a,b}$ of unique UVIT sources matched to PHAT (first 10 lines$^{c}$).}
\centering
\begin{tabular}{llllllllllllllllllll}
\topline
UVIT & x & y & RA & RAerr  & DEC & DECerr & F148W & F148Werr  \\
Field &  &  & (deg) & (deg) & (deg) & (deg) & (ABmag) & (ABmag)  \\
\hline
1 & 919.39 & 1418.74 & 10.72342 & 0.00006 & 41.27808 & 0.00014 & 23.25 & 0.19 \\
1 & 2234.93 & 935.09 & 10.70634 & 0.00013 & 41.29259 & 0.00015 & 23.26 & 0.19\\
1 & 745.05 & 2340.21 & 10.70040 & 0.00009 & 41.29403 & 0.00011 & 99.99 & 99.99\\
1 & 2217.53 & 1583.89 & 10.71760 & 0.00008 & 41.29573 & 0.00012 & 23.01 & 0.17\\
1 & 919.23 & 3867.61 & 10.73030 & 0.00012 & 41.28674 & 0.00007 & 23.95 & 0.25\\
1 & 2435.96 & 1405.44 & 10.71301 & 0.00021 & 41.29761 & 0.00017 & 99.99 & 99.99\\
1 & 2130.13 & 2629.24 & 10.71587 & 0.00016 & 41.26595 & 0.00008 & 99.99 & 99.99\\
1 & 2511.07 & 275.40 & 10.69291 & 0.00007 & 41.29296 & 0.00003 & 22.75 & 0.16\\
1 & 653.60 & 3727.83 & 10.73082 & 0.00022 & 41.29087 & 0.00005 & 23.84 & 0.24\\
1 & 2547.25 & 1728.91 & 10.71788 & 0.00014 & 41.30073 & 0.00014 & 23.12 & 0.18\\
\hline
F172M & F172Merr & F169M  & F169Merr & N219M & N219Merr & N279N & N279Nerr & UVIT-PHAT \\
(ABmag)  & (ABmag)   & (ABmag)  & (ABmag)  & (ABmag)  &(ABmag)   & (ABmag) & (ABmag)  & separation ($^{\prime\prime}$) \\
 \hline
23.35 & 0.29 & 22.44 & 0.14 & 99.99 & 99.99 & 21.25 & 0.22 & 0.172 \\
22.77 & 0.22 & 21.37 & 0.10 & 23.28 & 0.35 & 20.51 & 0.16 & 0.557\\
22.90 & 0.23 & 22.54 & 0.15 & 21.88 & 0.18 & 21.22 & 0.21 & 0.473\\
22.67 & 0.21 & 22.18 & 0.13 & 99.99 & 99.99 & 21.43 & 0.23 & 0.123\\
22.67 & 0.21 & 22.54 & 0.15 & 22.22 & 0.21 & 21.42 & 0.23 & 0.124\\
22.58 & 0.20 & 23.88 & 0.24 & 99.99 & 99.99 & 21.59 & 0.25 & 0.371\\
22.88 & 0.23 & 99.99 & 99.99 & 23.07 & 0.31 & 21.89 & 0.28 & 0.414\\
22.88 & 0.23 & 21.50 & 0.11 & 22.52 & 0.25 & 21.07 & 0.20 & 0.546\\
23.26 & 0.28 & 22.46 & 0.14 & 22.80 & 0.28 & 20.70 & 0.17 & 0.444\\
23.29 & 0.28 & 22.38 & 0.14 & 22.83 & 0.28 & 21.87 & 0.28 & 0.136\\

  \hline
\end{tabular}
\footnotesize
$\quad$\\
Notes: \\[0pt] 
a. The table is 18 columns by 712 rows. For display purposes here, the 18 columns are split into two sets of 9 columns. \\[0pt]
b. Magnitude and error values of 99.99 indicate no flux detected in that filter, blank values means that position not observed in that filter. \\[0pt]
c. The full table is available on-line in csv format as Table3\_Full.csv at github.com under repository denisleahy/M31-UVIT-sources-at-PHAT-positions. \\[0pt]
\label{table:posnerr} 
\end{table*}

To have useful photometry for CMDs, we removed those sources from the list with large errors ($\gtrsim0.3$) on the UVIT N279N magnitudes 
by restricting the N279N AB magnitude to be brighter than 22, leaving 851 sources in Field 2\footnote{the errors increase rapidly for magnitudes fainter than 22. E.g. at magnitude 23, which is roughly the detection limit, typical errors are $>0.45$ magnitudes.}.
The sky positions of these sources are shown in Figure 8.
We have made a catalogue of the sources from all four fields which were matched to unique PHAT sources 
($<1^{\prime\prime}$ separation, N279N magnitude$\le22$).
The catalogue contains the UVIT photometry for all available bands and consists of 1705 UVIT sources.
 A sample of part of the catalogue is given in Table 3 here, and
the full catalogue is available on line at https://github.com/denisleahy/M31-UVIT-sources-at-PHAT-positions.

\begin{figure}[!t]
\includegraphics[width=.99\columnwidth]{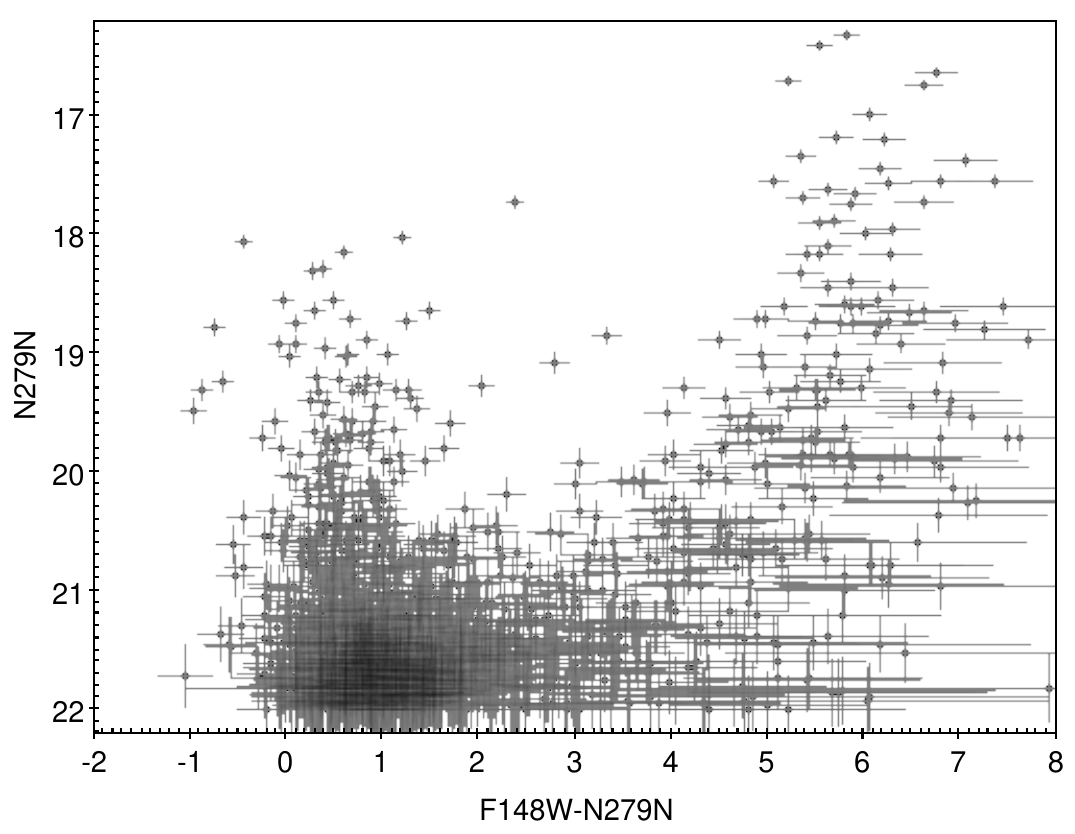}
\caption{UVIT FUV (150 nm F148W)-NUV (280nm N279N) color-magnitude diagram for the UVIT sources from Field 2 matched with PHAT sources.  Photometry errors are shown by the grey vertical and horizontal lines.}\label{fig5} 
\end{figure}


\begin{figure}[!t]
\includegraphics[width=.99\columnwidth]{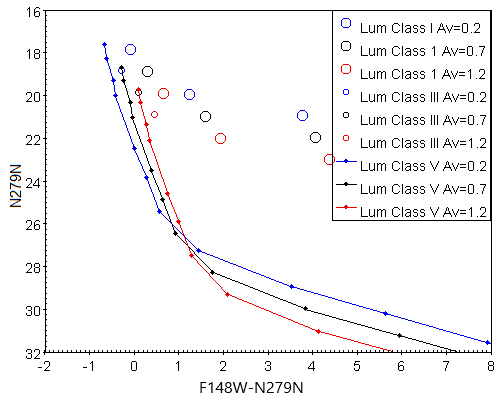}
\caption{UVIT FUV (150 nm F148W)-NUV (280nm N279N) color-magnitude diagram for a set of standard stars of luminosity
classes I, III and V. Those with extinction $A_V$=0.2 are marked in blue, with $A_V$=0.7  in black,
and $A_V$=1.2 in red.}\label{fig5}
\end{figure}

The UVIT CMD for these sources is shown in Figure 9, where the temperature indicator is  F148W$_{AB}$ magnitude - 
N279N$_{AB}$ magnitude (hotter stars on the left), and the luminosity indicator is the N279N$_{AB}$ magnitude.
The CMD shows two branches on the CMD: a near-vertical branch at F148W-N279N color $\simeq$0-1 and
 and a sloped branch to the right from color F148W-N279N $\simeq$2, N279N magnitude $\simeq$22 to 
 color F148W-N279N $\simeq$6, N279N magnitude $\simeq$18.
 The former likely corresponds to hot main-sequence
stars; and the latter may correspond to cooler evolved giant stars.

To verify this, a theoretical CMD was calculated for stars at the distance of M31, for main sequence stars of
spectral type O3V through to G0V, giants BOIII through to M0III and supergiants B0I through to M0I. The
radii, effective temperatures and log(g) values were taken from the recommended values for 
Castelli and Kurucz Atlas of Stellar Atmosphere Models at the
Space Telescope Science Institute  (https://www.stsci.edu/hst/instrumentation/reference-data-for-calibration-and-tools/astronomical-catalogs/castelli-and-kurucz-atlas). E.g. the O3V star has radius 15$R_{\odot}$, $T_{eff}$ =44800 K and log(g)=3.92, the G5III star has  radius 10$R_{\odot}$, $T_{eff}$ =51500 K and log(g)=2.54, and the
K5I star has 459$R_{\odot}$, $T_{eff}$ =3850 K and log(g)=0.00.

The theoretical CMD is shown in Figure 10 for 3 different values of extinction.  $A_V$=0.2 corresponds to the standard
foreground extinction from the Milky Way in the direction of M31. We have included  $A_V$=0.7 and  $A_V$=1.2,
corresponding to additional internal M31 extinction of 0.5 and 1.0. From Fig. 10, it is seen that normal upper
main sequence stars (O3V to B5V) in M31 fall in the same region as the observed UVIT sources. B0I, A0I, F0I and B0III stars 
also fall in the same area as observed UVIT sources.  None of the stars in the theoretical CMD fall in the region 
of the sloped branch on the right side of Fig. 9. However cooler main sequence stars (A0V through F5V) lie at the same 
color range (F148W-N279N color $\simeq$2-8) but much fainter N279N magnitudes. However foreground stars,
which are $\sim$100-1000 times closer, are brighter by 15 to 10 magnitudes, which places them in the same
N279N  magnitude range (16 to 21) as the observed UVIT sources. We conclude that most of the sources in
the upper-right quadrant of the observed UVIT CMD diagram are foreground stars.

With position matching of the UVIT sources to PHAT sources, we make the PHAT CMD 
for these same sources. This is shown in Figure 11, where the temperature indicator is 
F275W$_{Vega}$ magnitude  - 
F336W$_{Vega}$ magnitude, and the luminosity indicator is the F336W$_{Vega}$ magnitude.
The PHAT CMD shows two branches on the CMD: a near-vertical branch at color $\simeq$-0.5 and
 and a sloped branch to the right from color $\simeq$-0.5,  F336W magnitude $\simeq$21 to 
 color  $\simeq$2,  F336W magnitude $\simeq$17.
 
\begin{figure}[!t]
\includegraphics[width=.99\columnwidth]{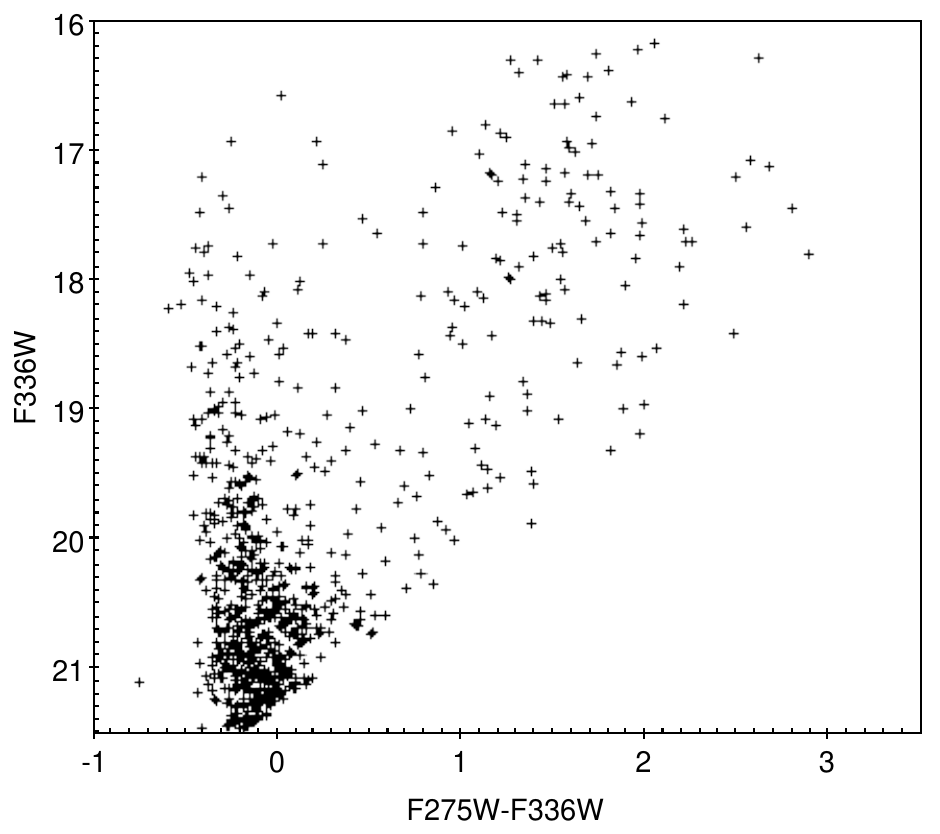}
\caption{PHAT NUV (275nm F275W)-optical (336 nm F336W) color-magnitude diagram for the UVIT sources from Field 2 matched with PHAT sources. Photometry errors are shown but are in most cases similar size to the plotted symbols.}\label{fig5}
\end{figure}

\begin{figure}[!t]
\includegraphics[width=.99\columnwidth]{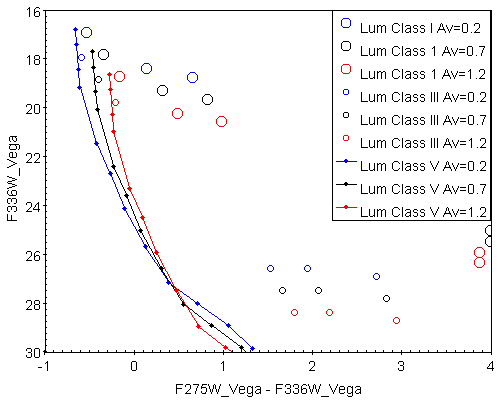}
\caption{PHAT NUV (F275W)-optical (F336W) color-magnitude diagram for a set of standard stars of luminosity
classes I, III and V. Those with extinction $A_V$=0.2 are marked in blue, with $A_V$=0.7  in black,
and $A_V$=1.2 in red.}\label{fig5}
\end{figure}

\begin{figure}[!t]
\includegraphics[width=.99\columnwidth]{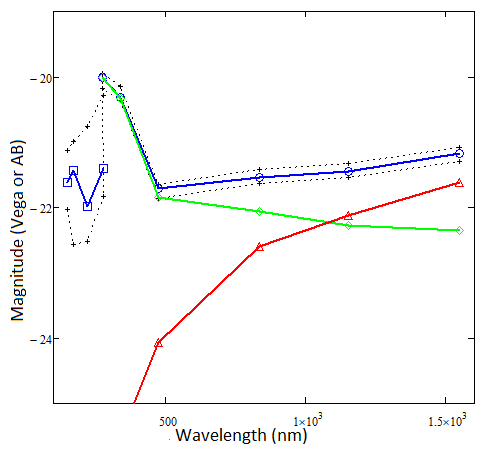}
\caption{Example stellar fit to SED from UVIT (F148W-150 nm, F172M-180 nm, N219M- 220 nm, N279N- 
280 nm) and  PHAT (F275W- 275 nm, F336W- 336 nm, F475W- 475 nm, F814W 814 nm, F110W- 1100 nm, 
F160W- 1600 nm).
The observed magnitudes and model magnitudes for UVIT are in the AB system;
 The observed magnitudes and model magnitudes for PHAT are in the Vega system. 
   The data upper and lower limits are the black dashed lines.
 The plotted total model (blue) is for a hot star (green) plus cool star (red).
 The fit parameters are: hot star T, radius and $A_V$ of 49,000 K, 6.6 $R_{sun}$ and 1.20; cool star 
T, radius and $A_V$ of 5700 K, 42 $R_{sun}$ and 1.56. 
}\label{fig5}
\end{figure}

 To examine the nature of these stars, the theoretical PHAT CMD was calculated for stars at the distance of M31,
 for the same parameters as for the theoretical UVIT CMD.
 The theoretical PHAT CMD is shown in Figure 12 for the 3 different values of extinction $A_V$=0.2, 0.7 and 1.2. 
 It is seen that normal upper
main sequence stars (O3V to B5V) in M31 fall in the same region of the PHAT CMD as the near vertical branch
of UVIT/PHAT sources.
As for the UVIT CMD, none of the cool stars are luminous enough to fall in the cool branch of the observed PHAT CMD.
However, foreground stars which are $\sim$100-1000 times closer (15 to10 magnitudes brighter) do fall in the 
same area as the cool branch of the observed  PHAT CMD.
Thus we find that the cool branch of both UVIT and PHAT CMDs is consistent with foreground stars, rather than 
stars in M31.

The UVIT FUV/NUV (4 filter bands) plus PHAT NUV-optical-IR (6 filter bands) photometry for a star from the main branch in
the UVIT CMD is shown in Figure 13. The drop between PHAT F275W (275 nm) band and the UVIT N279N (280 nm) band is 
a result of specifying  UVIT magnitudes in the AB system and PHAT magnitudes in the Vega system.
As noted above, for the 275-280 nm band the difference in the 2 systems is 1.48 mag, with Vega magnitudes brighter.
For Fig. 13 we plot the model spectrum for the UVIT bands in AB magnitudes and the model spectrum for the PHAT
bands in the Vega system, so the model and data show the same offsets between UVIT and PHAT bands.

A single star does not fit the data. This is caused by crowding in the longer optical and IR bands (see Leahy et al. 2017
for similar model fits for stars in the bulge of M31), so is not surprising.
The crowding is caused by two factors: the number of cool giant stars greatly exceeds that of hot main sequence stars,
caused by the combination of initial mass function and stellar lifetimes; the PSF for PHAT is larger for longer wavelengths.
A model consisting of a hot star plus a cool star fits the data well. The hot star is consistent with a massive young O-type star
($\sim$ 20 $M_{\odot}$); the cool star is consistent with a G-type evolved supergiant. 




\section{Conclusion}

The AstroSat Observatory has been carrying out a survey of M31 since 2017, which is nearly complete.
The primary goal is to obtain near ultraviolet and far ultraviolet observations with the UVIT
on AstroSat.
Surveying M31 requirs 19 fields each of $\sim$28 arcminute diameter.
All 19 fields were observed in the FUV F148W (150 nm) filter, and more than half of the fields 
observed in the NUV filters.
Recently we have developed new calibration and data processing methods which 
improve the astrometry and photometry of UVIT data.
With the new processing, the UVIT data has higher spatial resolution ($\simeq$1 arcsec) and 
better astrometry ($\simeq$0.2 arcsec).

This has allowed us to identify point sources detected by UVIT with sources at other wavelengths.
In particular, UVIT sources have been identified with stars observed at high resolution
with the Hubble Space Telescope as part of the Pan-chromatic Hubble
Andromeda Treasury project (PHAT, Williams et al. 2014). 
We are able to use color magnitude diagrams, using UVIT FUV-NUV colors and using 
PHAT NUV-visible colors to detect hot main sequence stars and to separate out likely foreground stars.
Future work will focus on carrying out UVIT photometry for the case of crowded UVIT sources, including multiple
PHAT stars near to eachother and star clusters, and on
modelling the multi-band UVIT-PHAT FUV through optical photometry of
different sets of stars in M31 to obtain good constraints on their properties.

\section*{Acknowledgements}
This project is undertaken with the financial support of the
Canadian Space Agency and of the Natural Sciences and
Engineering Research Council of Canada.
This publication uses the data from the AstroSat mission of the Indian Space Research Organisation (ISRO), archived at the Indian Space Science Data Centre (ISSDC). This publication uses UVIT data processed by the payload operations centre at IIA. The UVIT is built in collaboration between IIA, IUCAA, TIFR, ISRO and CSA. 
\vspace{-1em}



\end{document}